# Multiple transfer of angular momentum quanta from a spin-polarized hole to magnetic ions in ZnMnSe/ZnBeSe quantum wells

A. V. Akimov[1,2], A. V. Scherbakov[2], D. R. Yakovlev[1,2], I. A. Merkulov[2,3], M. Bayer[1], A. Waag[4], and L. W. Molenkamp[5]

[1] *Experimentelle Physik II, Universität Dortmund, D-44227 Dortmund, Germany*

[2] *A.F. Ioffe Physico-Technical Institute, 194021 St. Petersburg, Russia*

[3] *Oak Ridge National Laboratory, Oak Ridge, TN 37831-6016, USA*

[4] *Institute of Semiconductor Technology, Braunschweig Technical University, D-38106 Braunschweig, Germany*

[5] *Physikalisches Institut der Universität Würzburg, D-97074 Würzburg, Germany*

**Abstract**

The magnetization kinetics in (Zn,Mn)Se/(Zn,Be)Se quantum wells has been studied on a ps-time scale after pulsed laser excitation. The magnetization induced by an external magnetic field is reduced by up to 30% during ~100 ps due to spin and energy transfer from photocarriers to Mn spin system. The giant Zeeman splitting leads to a complete spin polarization of the carriers, resulting in a strong suppression of flip-flop processes between carriers and magnetic ions. Therefore a multiple angular momentum transfer from each spin-polarized hole to the Mn ions becomes the dominant mechanism in the magnetization dynamics. A model based on spin-momentum coupling in the valence band is suggested for explaining this transfer.

PACS: 75.75.+a, 72.25.Rb, 75.50.Pp, 78.47.+p

Currently great attention is directed towards the spin degrees of freedom from multiple fields, such as quantum computation or spintronics. However, manipulation of spins appears to be more challenging than manipulation of charges. Therefore it is crucial to develop a detailed understanding of interaction mechanisms between spin and other excitations in the solid, from which such a manipulation might be achieved. Spin and energy transfer from photoexcited carriers to magnetic ions in diluted magnetic semiconductors (DMS) is one of the key principles for ultrafast control of the magnetization in spintronic devices.

Studies of the coherent spin dynamics in transverse magnetic fields have shown fast (0.1-1 ps) spin transfer from polarized carriers to magnetic ions [1,2]. Further, experiments under continuous wave and nanosecond pulsed excitation have revealed a high efficiency of the energy transfer from carriers to the Mn ion spin system in longitudinal magnetic fields [3,4]. Energy and spin transfer are coupled to each other as the energy is conveyed from the carrier motional degree of freedom to the Zeeman-split Mn-spin reservoir. Thereby spin exchange occurs via a flip-flop mechanism, in which the spins of a carrier and a Mn ion flip in opposite directions conserving total spin [5]. At low temperatures this flip-flop mechanism should be suppressed in longitudinal magnetic fields, because the carrier spin subbands are well separated due to the giant Zeeman splitting and only the lowest spin subbands of electrons and holes are populated. To our knowledge, the dynamics of the carrier-magnetic ion interaction in this regime and in particular its efficiency has not been examined so far.

In this paper we report on the picosecond kinetics of the magnetic-field-induced magnetization in $Zn_{1-x}Mn_xSe$ quantum wells (QWs) under femtosecond optical excitation. We find a very efficient spin and energy transfer from the photoexcited holes to magnetic ions even for fully spin polarized holes. We explain this observation by mixing of the heavy- and light-hole states at finite wave vectors which allows for a multiple angular momentum transfer from each spin-polarized hole to the Mn ion system without change of the hole spin orientation.

The samples under study are nominally undoped $Zn_{1-x}Mn_xSe/Zn_{0.94}Be_{0.06}Se$ multiple quantum well structures grown by molecular beam epitaxy on (100) GaAs substrates [3]. Each sample contains five 10 nm wide $Zn_{1-x}Mn_xSe$ QWs separated by 20 nm $Zn_{0.94}Be_{0.06}Se$ barriers [3]. Four structures with different Mn concentrations $x$ = 0.005, 0.013, 0.03 and 0.1 were examined. Experiments were carried out at a bath temperature $T$ = 1.8 K and in magnetic fields up to 6 T applied along the structure growth axis (Faraday geometry). Time-resolved photoluminescence was chosen for experimental technique to investigate spin dynamics. Photocarriers were generated $Zn_{0.94}Be_{0.06}Se$ barriers by means of laser pulses from a frequency doubled 160 fs Ti-sapphire laser (photon energy 3.1 eV) with a repetition rate of 4 kHz. Energy per laser pulse was $P \sim 1$ nJ. Laser light was focused into a spot of 100 μm diameter [6]. After photogeneration carriers were collected in $Zn_{1-x}Mn_xSe$ QWs, loose their kinetic energy, form excitons and recombine. The photoluminescence (PL) was dispersed by a single 0.5-m monochromator and detected with a streak camera (time-resolution in this experiment was about 20 ps).

The giant Zeeman splitting of exciton states in DMS materials was used to monitor the dynamical changes of the magnetization [4]. In an external magnetic field $B$ the PL line is shifted to lower energies by an amount $\Delta E(B)$, which is proportional to the magnetization of the Mn ions $M(B)$ [7]:

$$\Delta E = \frac{1}{2}\frac{(\alpha-\beta)}{\mu_B g_{\mathrm{Mn}}} M(B) = \frac{1}{2}(\alpha-\beta) N_0 x S_{\mathrm{eff}} \mathrm{B}_{5/2}\left(\frac{5\mu_B g_{\mathrm{Mn}} B}{2k_B\left(T_{\mathrm{Mn}}+T_0\right)}\right). \quad (1)$$

Here $g_{\mathrm{Mn}} = 2$ is the $g$-factor of the $Mn^{2+}$ ground state. $S_{\mathrm{eff}}$ and $T_0$ are phenomenological parameters accounting for Mn-Mn interactions ($S_{\mathrm{eff}}$ = 2.5, 2.2, 1.8 and 1.0 as well as $T_0$ = 0.2, 0.55, 1.3, and 4 K for $x$ = 0.005, 0.013, 0.03 and 0.1, respectively [3]). $\mathrm{B}_{5/2}(z)$ is the modified Brillouin function. $N_0 = 10^{22}$ cm$^{-3}$ is the number of cations per unit volume. $N_0\alpha = 0.26$ eV and $N_0\beta = -1.31$ eV are the exchange constants for the conduction and valence (heavy hole) bands [8]. $T_{\mathrm{Mn}}$ is the Mn spin temperature (in thermal equilibrium $T_{\mathrm{Mn}} = T$). When $M$ decreases under a specific external impact this results in a decrease of $\Delta E$ by $\delta E$. Therefore $\delta E(t)/\Delta E = \Delta M(t)/M$ reflects the dynamics of relative magnetization changes $\Delta M(t)$. In the present work, the impact of the laser pulses is twofold: they disturb the equilibrium magnetization of the Mn system and they simultaneously permit optical detection of the magnetization dynamics.

Figure 1 shows two normalized spectral lines measured at delays of $t$ = 110 ps and 1500 ps after pulsed excitation in a sample with $x$ = 0.013 at $B$ = 2 T. The spectrum at later times is shifted to higher energies by $\delta E_{max}$ = 2 meV. This PL line shift occurs due to a dynamical decrease of magnetization. The left inset in Fig. 1 shows PL spectra measured for different delays under the same conditions. The shift of the line maxima as indicated by the dashed line is clearly seen. The PL line shift is also demonstrated in the right inset of Fig. 1 where the time evolution of the PL intensity $I(t)$ is shown at the energies corresponding to the maxima of the two PL lines in the main panel. The decay curves $I(t)$ cross each other at $t \sim 200$ ps. This can happen only when the PL line shifts in time to higher energies due to a decrease of the magnetization. No such spectral shift has been found at $B$ = 0 T.

To determine the PL peak energies, the spectra were analyzed by Gaussian line-shape fits, from which the time evolution $\delta E(t)/\Delta E = \Delta M(t)/M$ was obtained. $\Delta M(t)/M$ is plotted in Fig. 2 for three samples with different $x$. The $\Delta M$ = 0 value was taken from the PL line energy measured at low excitation pulse energy ($P$ = 0.06 nJ) where no dynamical shift $\delta E(t)$ was found. At small $t$ < 20 ps we observe a sharp rising edge whose details cannot be resolved with our setup. Later on $\Delta M(t)/M$ increases and then saturates for $t$ > 600 ps at $\Delta M_{\max}/M$. The rise time $\tau_M$ of $\Delta M(t)/M$ is comparable with the PL decay time $\tau_{\mathrm{PL}}$ = 160 ps (see dashed line in Fig. 2). Within the experimental error, the rise of $\Delta M(t)/M$ can be well described by the same time constant, as demonstrated by the solid lines in Fig. 2, which are single exponential fits using $\tau_M = \tau_{PL}$. Both times $\tau_M \approx \tau_{PL}$ are independent of magnetic field and decrease by 30% for an excitation energy decrease from $P$ = 1 to 0.06 nJ.

The dependence of $\Delta M_{max}/M$ on excitation energy $P$ for $x$ = 0.013 is shown in the inset of Fig. 2 for $B$ = 1 and 2 T. The experimental data follow approximately a linear dependence. The laser-induced changes of the magnetization are remarkably large (up to 30%). This change of magnetization corresponds to heating of the $Mn^{2+}$ ion spin system from 1.8 to 3.2 K under equilibrium conditions. The pulse energy dependence is almost 2 times weaker for $x$=0.03 than for $x$ = 0.013. No effect of the Mn heating is seen for $x$ = 0.1. The change of magnetization at fixed excitation pulse energy becomes smaller for increasing $B$, as seen in Fig. 3 for $x$ = 0.013. The temporal shift of the PL peak was also observed in a sample with the smallest Mn content $x$ = 0.005. The obtained results show that $\Delta M_{max}/M$ > 50% for $P$ = 1 nJ in a sample with $x$ = 0.005. However, due to very long spin-lattice relaxation time (~1 ms [4]) we could not observe a full recovery between pulses for a lowest possible repetition rate of the laser. This does not allow us to obtain the value for $\Delta M_{max}/M$ more accurately.

The decrease of magnetization $\Delta M(t)$ has to be associated with the dynamics of spin and energy transfer processes from photoexcited carriers to the Mn-ion spin system. We can exclude the influence of nonequilibrium phonons generated during energy relaxation and nonradiative recombination of carriers. These phonons are known to be important on a longer time scale and at higher excitation densities [4]. Therefore, the fast photoinduced magnetization changes are dominated by spin and energy transfer from hot photocarriers directly to the Mn ions. Due to slow spin-lattice relaxation of $Mn^{2+}$ ions [4] the transferred energy and spin accumulate in the Mn system. As a result, $\Delta M(t)$ increases continuously with time, as long as photoexcited carriers are present. It saturates at a level $\Delta M_{max}$ for times $t > \tau_M \sim \tau_{PL}$. We neglect in our consideration effects related to exciton formation and specifics of exciton interaction with magnetic ions. Excess kinetic energy of photoexcited carriers and typical values for the giant

Zeeman splitting exceed the exciton binding energy. Therefore, the photocarriers relax separately to the lowest spin sublevels and only there they are bound into excitons.

Let us discuss what transfer mechanism can provide the huge reduction of magnetization $\Delta M_{\max}$ within the short time $\tau_M$. Our analysis will be based on the theoretical approach developed in Ref. [5]. A single spin flip-flop scattering event between carrier and ion is faster than a few picoseconds [1, 2], but the energy shuffled to the Mn ion is rather small, e.g. $\mu_B g_{\mathrm{Mn}} B$ = 0.23 meV at $B$ = 2 T. Transfer of a significantly larger amount of energy requires multiple scattering, which in principle is possible within the typical carrier lifetime of 100 ps. However, after the first flip-flop the carrier has an unfavourable spin orientation and can no longer transfer energy to Mn ions until its spin relaxes back by any mechanism except exchange scattering with Mn. The latter would reverse the effect of the original flip-flop process and, therefore, cannot be considered as contribution to multiple spin transfer, as it would transfer spin and energy back from a Mn ion to a carrier.

The maximum relative change of the magnetization can be written as:

$$\frac{\Delta M_{\max}}{M}=\frac{\lambda_{e-\mathrm{Mn}}n_e+\lambda_{h-\mathrm{Mn}}n_h}{xdN_0S_{\mathrm{eff}}B_{5/2}\left[\dfrac{5\mu_B g_{\mathrm{Mn}}B}{2k_B\left(T+T_0\right)}\right]}, \qquad (2)$$

where $n_{e,h}$ are the electron ($e$) and hole ($h$) densities generated by a laser pulse in a QW of width $d$ = 10 nm. $\lambda_{e-\mathrm{Mn}}$ ($\lambda_{h-\mathrm{Mn}}$) is the number of irreversible exchange scattering events with Mn for a photoexcited electron (hole). Eq. (2) allows us to estimate whether the multiple spin transfer ($\lambda_{e-\mathrm{Mn}}+\lambda_{h-\mathrm{Mn}} >> 1$) is required to explain the experimental results.

Let us first analyze the electron contribution. The electron spin relaxation time in nonmagnetic QWs varies between 100 ps and a few ns [9,10] and is about equal or longer than the carrier lifetime $\tau_{\mathrm{PL}}$ ~ 150 ps in our experiments. Therefore $\lambda_{e-\mathrm{Mn}}$ ~ 1, so that only a single transfer event on average can take place within $\tau_M \sim \tau_{\mathrm{PL}}$. We estimate the carrier density for $P$ = 1 nJ excitation energy as $n_{e,h}$ = (3÷7)×$10^{11}$ $\mathrm{cm}^{-2}$. This value is much less than the sheet density of Mn ions in a QW ($1.3\times10^{14}$ cm$^{-2}$ for $x$ = 0.013). From Eq. (2) with $x$ = 0.013, $B$ = 2 T, $\lambda_{e-\mathrm{Mn}}$ = 1 and $n_e$ = $5\times10^{11}$ cm$^{-2}$ we find a magnetization change of $\Delta M_{\max}/M$ ~ 0.003, which is almost two orders of magnitude smaller than the experimentally measured value of 0.12. Therefore, the electrons only cannot explain the large change in magnetization.

Contrary to electrons, the spin-relaxation time of holes is very fast ($\tau_S^h$ = 0.1÷1 ps), even in nonmagnetic semiconductors. Due to the strong spin-orbit coupling any momentum scattering of a hole likely changes also its spin state [11]. Therefore, holes may undergo multiple spin-flip transitions. However, this is valid for holes whose kinetic energy exceeds the Zeeman splitting. For our experimental conditions this is the case for the first 50 ps after laser excitation, when carriers relax energetically from the barriers into the QWs. By analyzing the high-energy slope of the PL spectra (Fig. 4) at $B$ = 0 T for different delays ($P$ = 1 nJ) the hole temperature at $t$ = 80 ps can be estimated by $T_h$ = 32 K only. $T_h$ decreases with time and for $t$ > 500 ps the high energy wing is identical to the one at low excitation. As one can see from the diagram in Fig. 4 the energy difference between the $\left|+3/2\right\rangle$ and $\left|+1/2\right\rangle$ hole spin states is $E_0$ = 22 meV ($x$ = 0.013 at $B$ = 2 T). At $T_h$ = 32 K, the population of the upper $\left|+1/2\right\rangle$ light-hole subband is by a factor $\exp(-E_0/k_B T_h)$=$6\times10^{-4}$ smaller than that of the low energy $\left|+3/2\right\rangle$ subband. The low $\left|+1/2\right\rangle$ population is confirmed also by the absence of a PL signal related to the light holes. Thus due to the very low population of the high energy states, not enough flip-flops are provided by hot holes, even for multiple hole scattering: The magnetization changes for $t \geq$ 80 ps cannot exceed $\Delta M_{\max}/M$ $\leq$ 0.003, also by far too small to account for the experimental value $\Delta M_{\max}/M$ = 0.12 in Fig. 2, especially because about 70% of the demagnetization occurs for $t \geq$ 80 ps.

Therefore another mechanism has to be considered, which can provide spin and energy transfer from spin-polarized holes to Mn ions, while the initial and the final states of the holes belong to the same spin subband $\left|+3/2\right\rangle$. Such, from first sight very surprising, scattering of a hole at the contact potential of Mn is possible due to the $(\mathbf{k}\mathbf{J})^2$ term with the 3D hole spin operator $\mathbf{J}$ ($j$=3/2) in the Luttinger Hamiltonian [12]. Due to this coupling, at finite $k$ the heavy hole state $|+3/2\rangle$ mixes with the light hole state $|+1/2\rangle$, resulting in non-zero matrix elements between even and odd subbands of hole quantization in quantum wells. In lowest order perturbation theory the scattering rate for a heavy hole $|+3/2\rangle$ at a Mn ion is proportional to the ratio of the hole kinetic energy to the splitting $E_1$ between the first heavy-hole and the second light-hole subbands. For a nondegenerate hole gas this ratio is proportional to $k_B T_h / E_1$.

An explicit calculation of the hole relaxation rate within the lowest heavy-hole subband requires an extended numerical computation, which is beyond the scope of this paper. Qualitatively, the probability of elementary flip-flop transitions $|\pm 3/2\rangle \leftrightarrow |\pm 1/2\rangle$ is higher than that for transitions within the $|+3/2\rangle$ hole subband. However, the latter scattering channel does not require any population of levels with energies $E >> k_B T_h$. In the sample with $x$ = 0.013 Mn-content $E_1 \sim 40$ meV at $B$ = 2T. The relatively small parameter $k_B T_h / E_1 = 0.07$ in our first order perturbation analysis still overpowers the even smaller population $\exp(-E_0 / k_B T_h) < 6 \times 10^{-4}$, which governs the flip-flop transition efficiency. Thus multiple spin and energy transfer is not restricted to flip-flop transitions only but may continue during the entire relaxation cascade. This scenario may explain our experimental observation.

Photoexicted hot carriers relax mostly via emission of optical phonons until their kinetic energy is higher than $\hbar\omega_{LO}$. Further the relaxation energy is shared between $Mn^{2+}$ ions and acoustic phonons. For semiquantitative estimation we introduce a parameter $\chi = \lambda_{h-\mathrm{ph}} / \lambda_{h-\mathrm{Mn}}$, where $\lambda_{h-\mathrm{ph}}$ is the mean number of acoustic phonons with average energy $\hbar\omega \sim 1$ meV [13] emitted by a hole on a relaxation path with energy $E_{Rh}$. The parameter $\chi$ describes also the ratio of the mean times for transitions $h$-Mn, when a quantum $\mu_B g_{Mn} B$ is transferred to $Mn^{2+}$ ion and $h$-$ph$ when an acoustic phonon is emitted $\chi = \tau_{h-\mathrm{ph}} / \tau_{h-\mathrm{Mn}}$ . It is easy to show that

$$E_{Rh} = \lambda_{h-Mn} (\chi \hbar\omega + \mu_B g_{Mn} B) \qquad (3),$$

We estimate the value for $E_{Rh}$ assuming that the energy transfer from a hole to Mn ion is efficient only when its kinetic energy becomes smaller than the energy of optical phonons $\hbar\omega_{LO}$ = 31 meV. We also take into account that the electrons can transfer their kinetic energy to the holes as a result of electron-hole scattering and therefore $E_{Rh}$ increases by about a factor of two. Thus taking the values $E_{Rh}$ = 62 meV, , evaluating $\lambda_{h-\mathrm{Mn}}$ from Eq. (3) and substituting it in Eq. (2) with $\lambda_{e-\mathrm{Mn}} = 1$ it is possible to estimate the values of $\Delta M_{\max}/M$ for various $\chi$, $x$ and $B$. Curves in Fig. 3 are the calculated magnetic field dependencies of $\Delta M_{\max}/M$ for the sample with $x$ = 0.013 and different parameters $\chi$. The best agreement with experimental data is recieved for $\chi$ = 0.4. Applying the same procedure for the sample with $x$ = 0.03 we get the best agreement for $\chi$ = 0.8. The value of $\chi$ is even larger in the sample with $x$ = 0.1 where the experimental noise does not allow us to measure the value for $\Delta M_{\max}/M$ precisely. For the sample with low Mn content, $x$ = 0.005 we may conclude that $\chi$ ~ 0.3. Thus, qualitatively $\chi$ increases with the increase of $x$, which means that the competition between the scattering of holes on Mn ions and acoustic phonons goes in favor of relaxation on phonons with the increase of Mn content. Correspondingly $\lambda_{h-\mathrm{Mn}}$ decreases with the increase of $x$. The reason for this may be the relative increase of Mn clusters which are not active in spin transfer process from carriers to isolated Mn ions, which finally determine the magnetization. It is well known that the role of clusters in spin dynamics becomes important already at very low Mn contents ($x$ ~ 0.001) [14, 15].

The calculated curves in Fig.3 show the rapid increase of $\Delta M_{\max}/M$ at low magnetic fields. The experimentally observed increase of $\Delta M_{\max}/M$ at $B$ = 1 T is almost twice less than predicted by the model with $\chi$ = 0.4. Apparently this is due to the fact that the spin system is already heated in equilibrium ($T$ = 1.8 K). This is important at low $B$ when $k_B T > \mu_B g_{Mn} B$ and may lead to the essential decrease of $\Delta M_{\max}/M$ in comparison with the qualitative model, which includes only spin-flip with the increase of a spin projection in Mn ions.

To conclude, in DMS quantum wells hot carriers generated by femtosecond laser pulses can efficiently transfer their excess kinetic energy to magnetic ions on a time scale of ~100 ps and induce changes of the magnetization up to 30%. The analysis shows that a flip-flop mechanism is insufficient to describe the data. A mechanism that can provide such a large effect through scattering of heavy-holes within one Zeeman subband at Mn ions is suggested.

**Acknowledgements**. We acknowledge financial support by the BMBF (nanoquit), by DARPA, by the RFBR 05-02-16389-a and the Russian Academy of Sciences. AVA has been granted by the Deutsche Forschungsgemeinschaft through the Mercator guest professorship programme. Research stays of AVS in Dortmund were funded by the Deutsche Forschungsgemeinschaft via grants 436RUS17/32/04 and 436RUS17/16/05.

**Figure captions**

Fig. 1. Normalized PL spectra of $Zn_{0.987}Mn_{0.013}Se/Zn_{0.94}Be_{0.06}Se$ QWs measured at delay times $t$ = 110 (dashed line) and 1500 ps (solid line) integrated over time intervals of 10 and 220 ps, respectively. The left inset shows normalized photoluminescence spectra measured at different delays in a time window of 80 ps. The dashed line helps to follow the spectral shift of the PL maxima. The right inset shows the time evolution $I(t)$ of PL measured at the PL maxima of the spectra in the main panel.

Fig. 2. Time evolution of $\Delta M(t)/M$ in two samples. Solid lines are single exponential fits to the experimental data with $\tau_M = \tau_{PL} = 160$ ps ($x$ = 0.013) and 150 ps ($x$ = 0.03). Dashed line is the time evolution of PL intensity $I(t)$ integrated over a spectral range of 15 meV for the $x$ = 0.013 sample. An inset shows $\Delta M_{\max}/M$ as a function of excitation energy for two values of $B$ for the sample with $x$ = 0.013. Solid line serves as a guide to the eye for a linear dependence on $P$.

Fig. 3. Magnetic field dependence of $\Delta M_{\max}/M$, measured in the sample with $x$ = 0.013 for $P$ = 0.7 nJ. Dependencies of $\Delta M_{\max}/M$ on $B$ for different fitting parameters $\chi$ are given by lines (see text for details). The inset is a scheme of transitions involved in multiple spin transfer from carriers to $Mn^{2+}$ ions. There are only flip-flop transitions for electrons, while for holes the transfer of spin may be possible during the relaxation in a single Zeeman subband $|+3/2\rangle$.

Fig. 4. Upper panel: energy diagram of the studied QW structure ($x$ = 0.013) demonstrating the giant Zeeman splitting. Lower panel: normalized PL spectra measured at different delays for $P$ = 1 nJ (solid lines) and 0.06 nJ (dashed line). The doublet structure corresponds to a neutral $X$ and charged $X^-$ excitons. The latest quenches at $B \geq 1$ T [3].

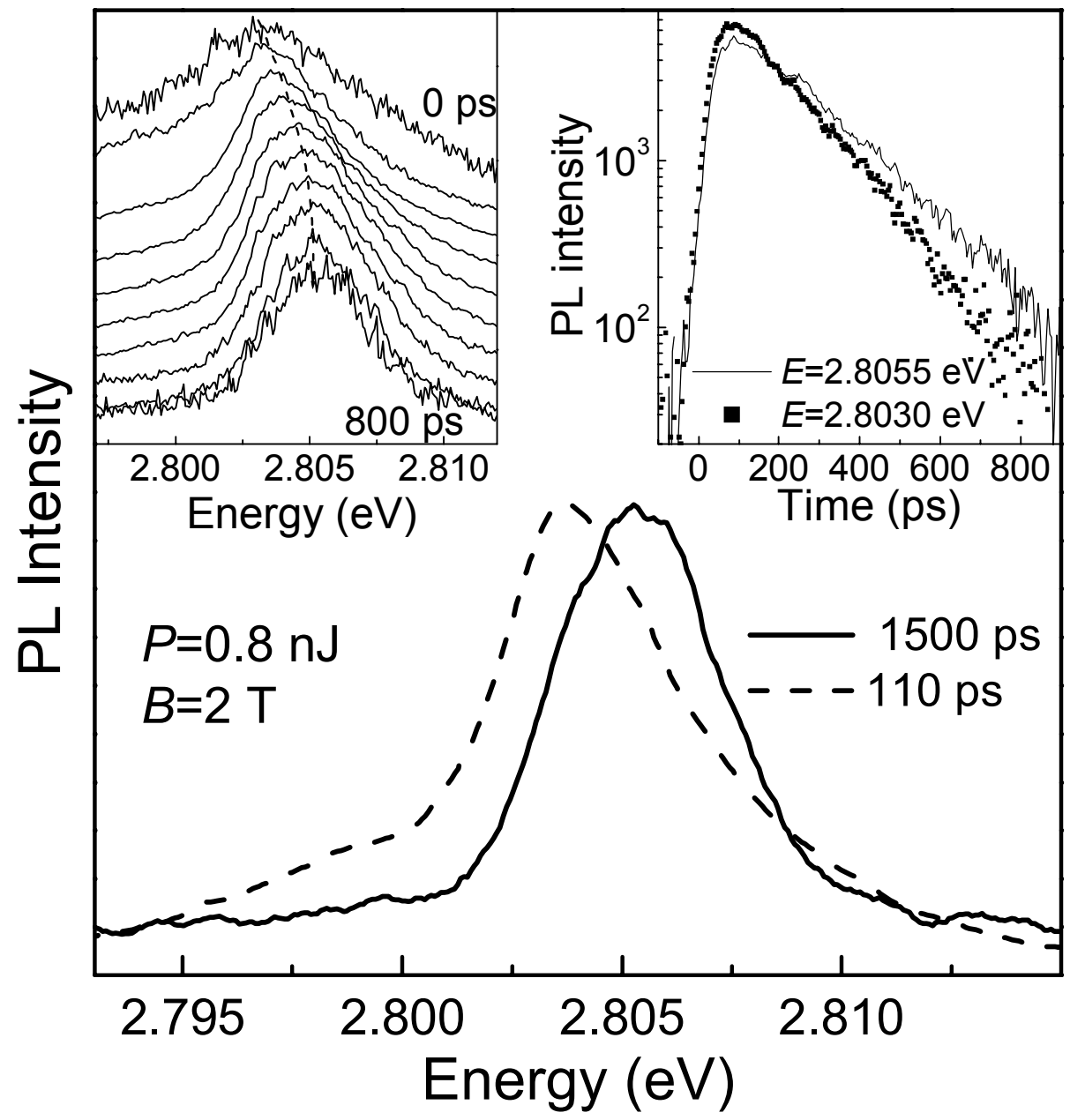

Fig. 1. A. V. Akimov et al. "*Multiple transfer of angular momentum…*"

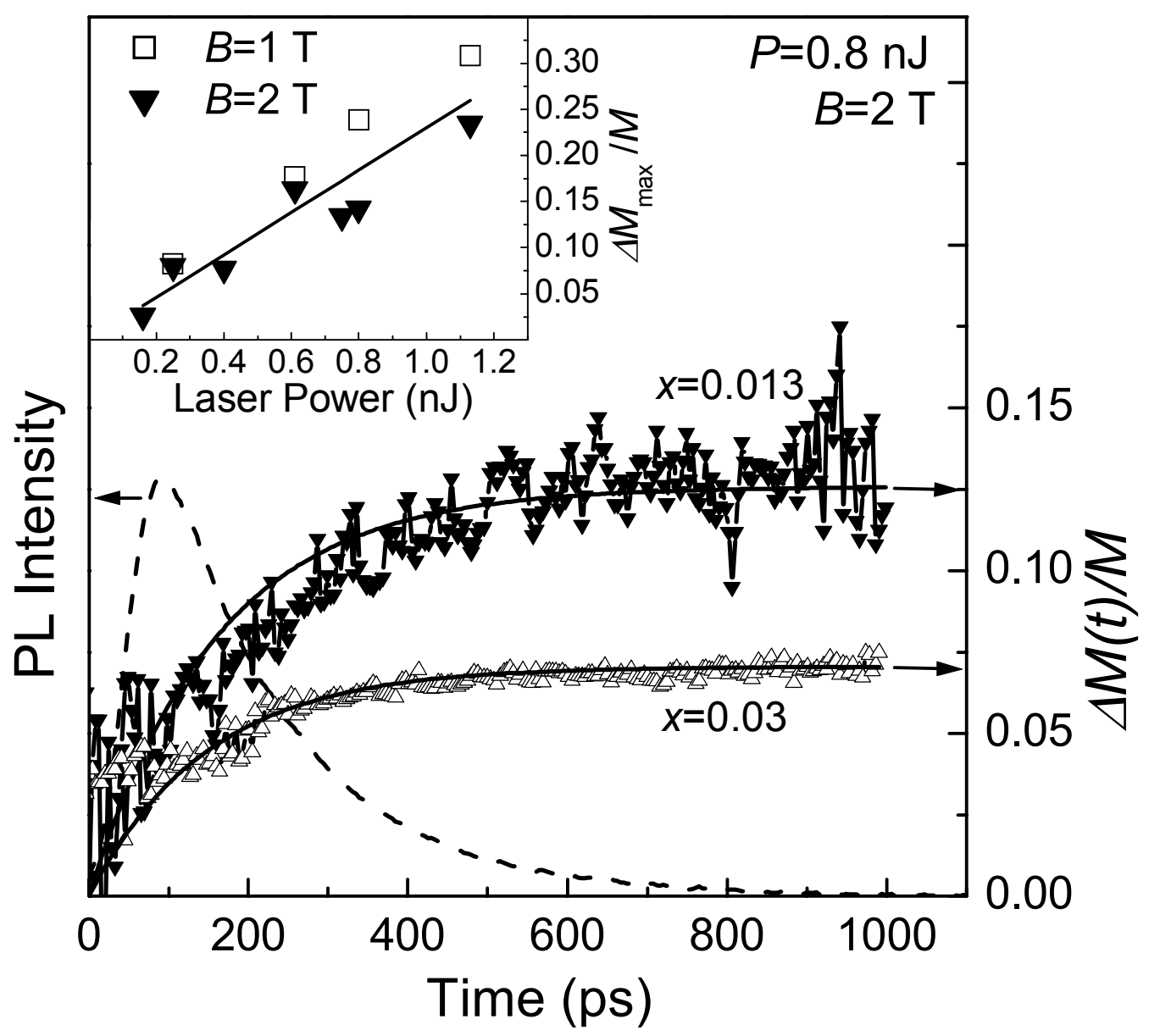

Fig. 2. A. V. Akimov et al. "*Multiple transfer of angular momentum…*"

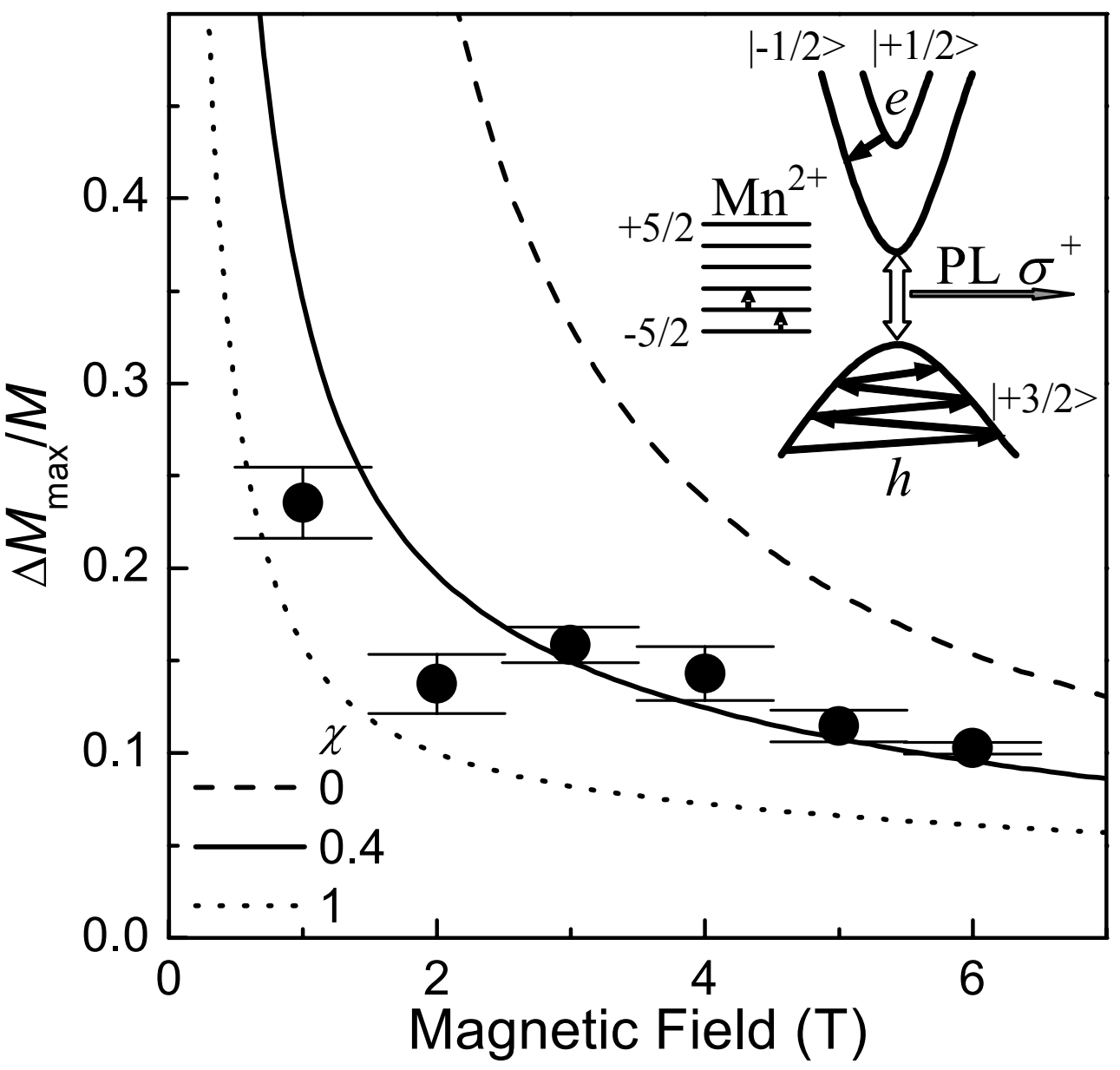

Fig. 3. A. V. Akimov et al. "*Multiple transfer of angular momentum…*"

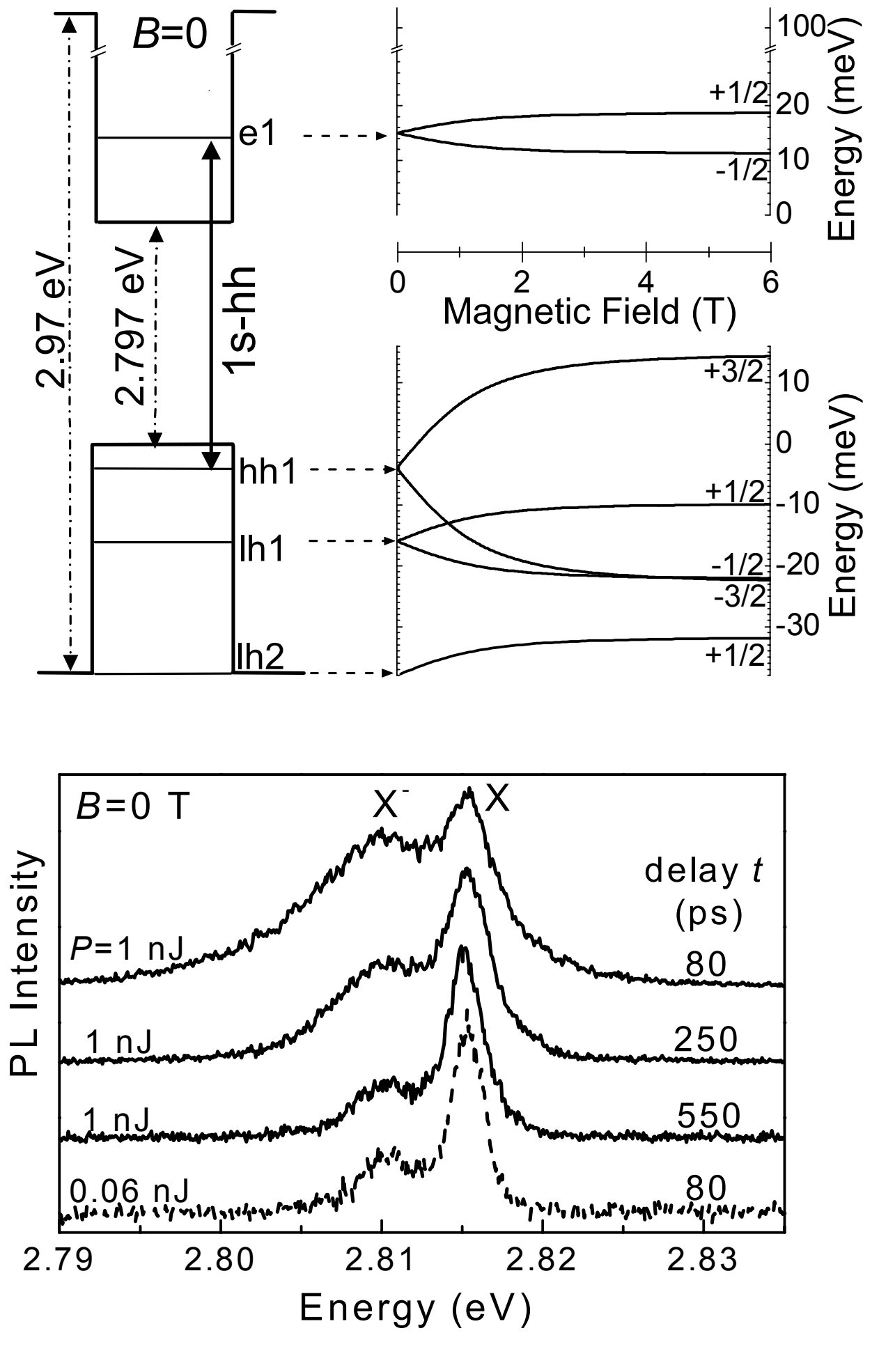

Fig. 4. A. V. Akimov et al. "*Multiple transfer of angular momentum…*"